%% file: manuscript.tex
\documentclass[aps, prl, twocolumn, superscriptaddress]{revtex4-1}

\usepackage{amsmath}
\usepackage{amssymb}
\usepackage{mathrsfs}
\usepackage{bm}
\usepackage{braket}

\usepackage{cases}

\usepackage{mathtools}
\newcommand{\defeq}{\vcentcolon=}

\usepackage[colorlinks=true]{hyperref}

\usepackage{comment}

\def \ux {\vec{u}_x}
\def \uy {\vec{u}_y}
\def \uz {\vec{u}_z}

\begin{document}

\title{Electron spin transport driven by surface plasmon polariton}

\author{Daigo Oue}
\affiliation{Kavli Institute for Theoretical Sciences, University of Chinese Academy of Sciences, Beijing, 100190, China.}
\affiliation{The Blackett Laboratory, Department of Physics, Imperial College London, Prince Consort Road, Kensington, London SW7 2AZ, United Kingdom}
\author{Mamoru Matsuo}
\affiliation{Kavli Institute for Theoretical Sciences, University of Chinese Academy of Sciences, Beijing, 100190, China.}
\affiliation{CAS Center for Excellence in Topological Quantum Computation, University of Chinese Academy of Sciences, Beijing 100190, China}

\date{\today}

\begin{abstract}
  We propose a mechanism of angular momentum conversion from optical transverse spin in surface plasmon polaritons (SPPs) to conduction electron spin.
  Free electrons in the metal follow the transversally spinning electric field of SPP,
  and the resulting orbital motions create inhomogeneous static magnetisation in the metal.
  By solving the spin diffusion equation in the SPP,
  we find that the magnetisation field generates an electron spin current.
  We show that there exists a resonant condition where the spin current is resonantly enhanced, and the polarisation of the spin current is flipped.
  Our theory reveals a novel functionality of SPP as a spin current source.
\end{abstract}

\maketitle

\paragraph{Introduction.---}
\label{para:Introduction---}
The optical transverse spin is one of the universal properties of evanescent waves \cite{van2016universal, fang2017intrinsic, oue2019dissipation}.
It is an exotic circular polarisation of evanescent field whose spinning direction of the field is perpendicular to the propagation direction unlike ordinary propagating fields.
When the decay direction of the evanescent field is not parallel to its propagation direction,
the transverse spin exists in the evanescent fields due to the transversality requirement from Gauss law.

\par
As surface plasmon polariton (SPP) is an electromagnetic wave coupled with plasma oscillations localised at a metal-dielectric interface \cite{novotny2012principles}.
The SPP possesses transverse spin \cite{bliokh2012transverse, bliokh2015quantum} because the decay direction is normal to the interface along which the SPP propagates.
The transverse spin in the SPP generates inhomogeneous magnetisation field in the metal.
This is because the electron gas in the metal makes orbital motions,
following the transversally spinning electric field of the SPP.
The electric current given by the curl of this magnetisation is divergenceless ($\nabla \cdot \nabla \times = 0$),
and it cannot be detected \cite{bliokh2017optical_new_j_phys}.
However, a detectable spin current is generated by the magnetisation as shown below.

\par
In metals, there are generally two kinds of electronic transport, not only charge currents but also spin currents.
It is known that the spin transport is driven in media with the presence of spin dependent potentials,
such as a strong spin-orbit coupling \cite{kato2004observation, wunderlich2005experimental, kimura2007room, kohda2012spin},
and spin-vorticity coupling \cite{takahashi2016spin, kobayashi2017spin, okano2019nonreciprocal}.
In particular,
the gradient of effective magnetic fields is utilised in \cite{kohda2012spin, takahashi2016spin, kobayashi2017spin, okano2019nonreciprocal}.
The effective magnetic fields are created by the inhomogeneous spin-orbit coupling \cite{kohda2012spin} or by spin-vorticity coupling \cite{takahashi2016spin, kobayashi2017spin, okano2019nonreciprocal}.
That is, a variety of the Stern-Gerlach-like effects are exploited for generating spin currents. 
In this work, we identify the inhomogeneous magnetisation field of SPPs as a new candidate for driving spin currents,
and thus the transverse spin in SPP could be detected via spin current measurements.
However, to the best of our knowledge, there have not been any experiments or theories related to the spin transport mediated by SPP.

\par
\begin{figure}[htbp]
  \centering
  \includegraphics[width=\linewidth]{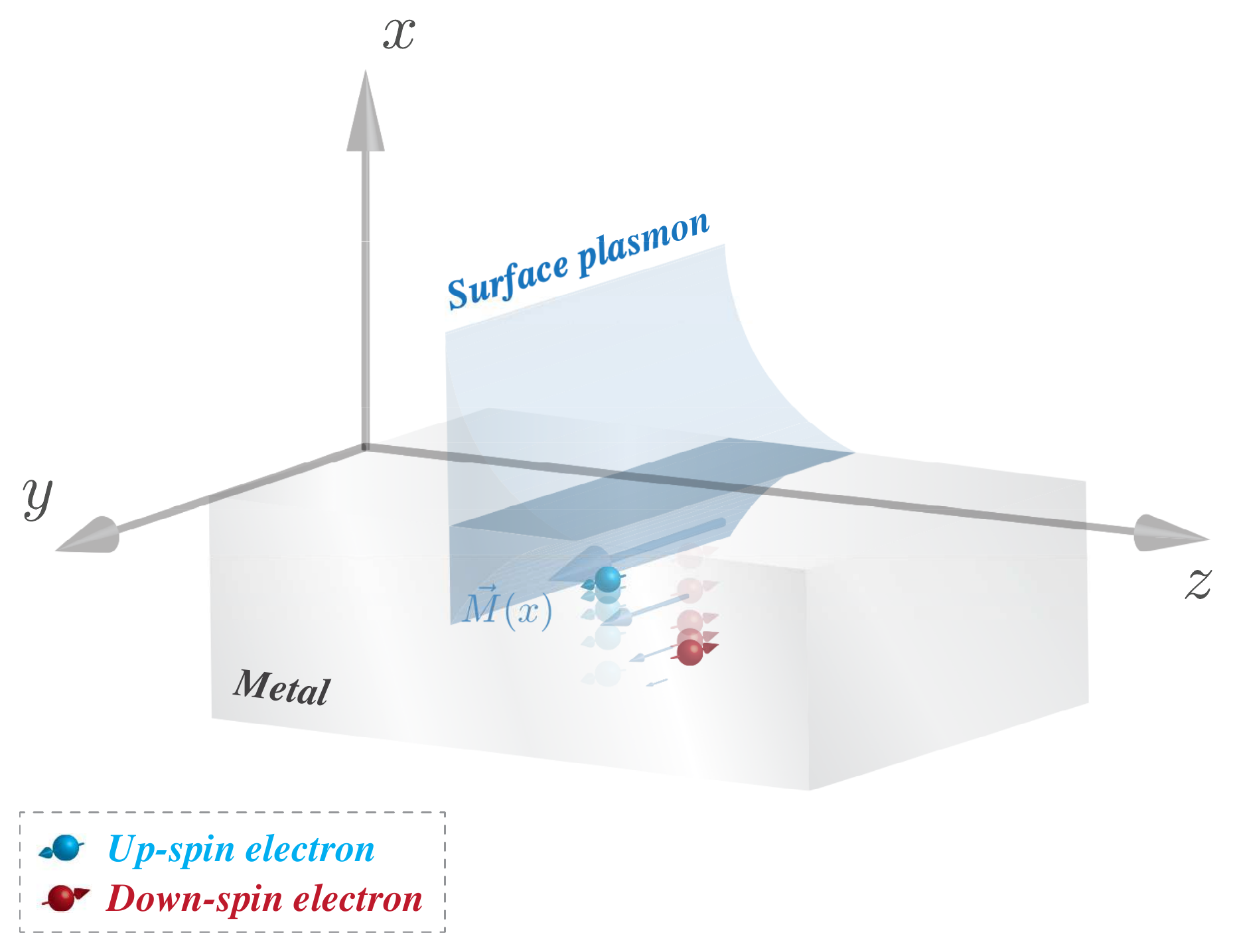}
  \caption{
    Schematic of a setup for angular momentum conversion from SPP to electron spin, 
    which we analyse in this paper.
    We have a dielectric-metal interface where a SPP is excited.
    The transverse spin of SPP excites the orbital motion of electrons,
    which create the magnetisation field in the metal.
    Since the transverse spin density decays exponentially, 
    there is steep gradient of the magnetisation field in the metal.
    This inhomogeneous magnetic field drives the spin current carried by conduction electrons in the metal,
    whose flow direction is perpendicular to the interface.
  }
  \label{fig:am_conversion}
\end{figure}

In this paper, we solve the spin diffusion equation in the presence of inhomogeneous magnetisation generated by SPP (FIG. \ref{fig:am_conversion}).
and we find that the spin accumulation and thus the diffusive spin current are created by the inhomogeneous magnetisation.
The spin current can be detected since the divergence of the spin flow does not vanish unlike the charge current.
This means that the transverse spin in SPP drives the electron spin current in the metal.
We use Gaussian units in this paper except in the final part where we estimate the order of the mangitude of the spin currents so as to investigate whether they are measurable or not.
We bridge two seemingly distant fields: plasmonics and spintronics.

\paragraph{Transverse spin and inhomogeneous magnetisation in a surface plasmon polariton.---}
\label{para:Transverse spin & inhomo. magnetisation---}
The electric and magnetic field of a SPP are given by \cite{nkoma1974elementary, zayats2005nano, bliokh2012transverse, bliokh2017optical_new_j_phys}
\begin{align}
  \vec{E} &= E_0 \left[ \left( \ux - i\frac{\kappa_1}{k_p} \uz \right) e^{-\kappa_1 x} \theta(x) \right. \notag \\
          &\hspace{1.9em} + \left. \epsilon^{-1} \left( \ux + i\frac{\kappa_2}{k_p} \uz \right) e^{\kappa_2 x} \theta(-x)\right] e^{i k_p z}, \label{eq:E_SPP} \\
  \vec{H} &= E_0 \frac{k_0}{k_p}\uy \left[ e^{-\kappa_1 x} \theta(x) + e^{\kappa_2 x} \theta(-x) \right] e^{i k_p z}. \label{eq:H_SPP} 
\end{align}
Here we use Heaviside unit step function $\theta(x)$,
and set $k_0 = \omega/c$.
The wavenumber of the SPP is defined by
\begin{align}
  k_p &= \frac{\sqrt{-\epsilon}k_0}{\sqrt{-1-\epsilon}},
  \label{eq:k_SPP}
\end{align}
and the decay coefficients in vacuum and in metal are defined by
\begin{align}
  \kappa_1 &= \frac{k_0}{\sqrt{-1-\epsilon}},
  \label{eq:decay_vac}
  \\
  \kappa_2 &= \frac{-\epsilon k_0}{\sqrt{-1-\epsilon}},
  \label{eq:decay_metal}
\end{align}
respectively.
We can obtain these quantities by 
applying the boundary matching condition at the interface to Maxwell's equations with Drude free electron model 
\begin{align}
  \epsilon(\omega) = 1 - \frac{\omega_p^2}{\omega(\omega+i\gamma)}.
  \label{eq:Drude_model}
\end{align}
Here, we set the plasma frequency $\omega_p^2 = 4\pi n e^2/m$ and the permeability $\mu = 1$.
In the equation (\ref{eq:E_SPP}), we can see the imaginary unit $i$ at $\uz$ while not at $\ux$.
This implies that there is phase difference between the longitudinal $z$ component and the transverse $x$ component of the field,
and the electric field rotates in the transverse $y$ direction both on the dielectric side and on the metal side.
Note that the rotation direction in the dielectric side and that in the metal side are opposite to each other.

\par We use the Minkowski representation for the spin angular momentum density of an electromagnetic field,
\begin{align}
  \vec{S} \defeq \frac{g}{2} \mathfrak{Im} \left( \tilde{\epsilon} \vec{E}^* \times \vec{E} + \tilde{\mu} \vec{H}^* \times \vec{H} \right).
  \label{eq:spinAM}
\end{align}
Here $g = (8 \pi \omega)^{-1}$ is a Gaussian unit factor,
the group permittivity $\tilde{\epsilon} = \frac{\mathrm{d} (\omega \epsilon)}{\mathrm{d}\omega}$,
and permeability $\tilde{\mu} = \frac{\mathrm{d} (\omega \mu)}{\mathrm{d}\omega}$.
As Bliokh \textit{et al.} demonstrated in the literature \cite{bliokh2017optical_new_j_phys},
we can decompose the Minkowski representation of the spin angular momentum density of a SPP into two contributions.
One is a contribution from the electromagnetic field and the other is from the kinetic motion of electrons in the metal,
which corresponds to the dispersion corrected term of the spin angular momentum density.
For a SPP, we have
\begin{align}
  \vec{S} &= \vec{S}_{em} + \vec{S}_{mat},\notag \\
          &= \frac{g\epsilon}{2} \mathfrak{Im} \left( \vec{E}^* \times \vec{E} \right) + \frac{g \omega}{2} \frac{\mathrm{d} \epsilon}{\mathrm{d} \omega} \mathfrak{Im} \left( \vec{E}^* \times \vec{E} \right).
          \label{eq:decomposition_spinAM}
\end{align}
Here we ignore the magnetic field contribution to the spin angular momentum density,
because there is no rotation and $\mathfrak{Im} \left( \vec{H}^* \times \vec{H} \right) = 0$ in SPPs.

\par
\begin{figure}[htbp]
  \centering
  \includegraphics[width=\linewidth]{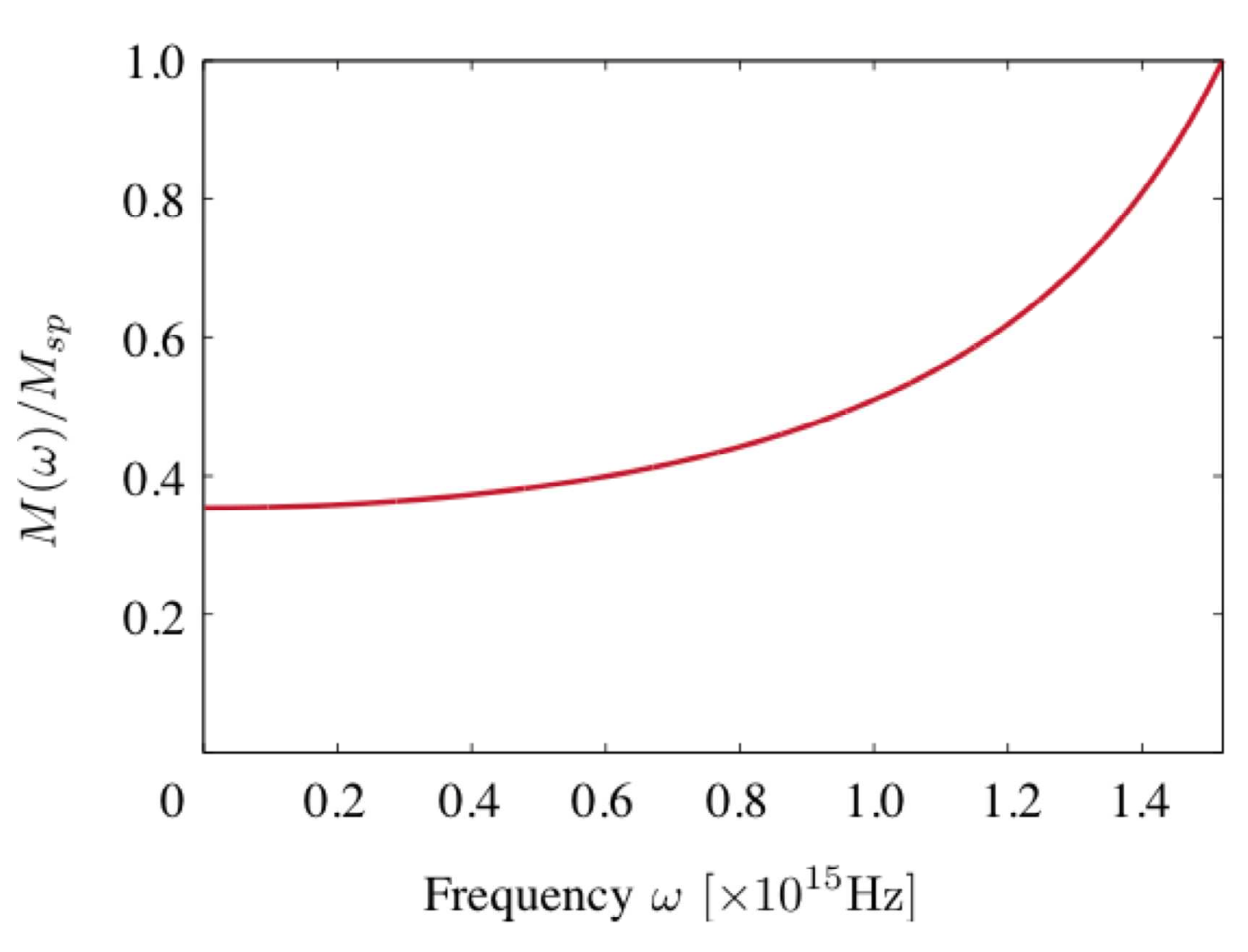}
  \caption
  {
    Frequency dependence of the magnetisation density in a SPP.
    Here we define $M_{sp} \equiv M(\omega_{sp})$.
    It is clear that the magnetisation is a monotonically increasing function of the SPP frequency.
    We use Drude parameter of gold $\omega_p = 2.15 \times 10^{15}\ \mathrm{Hz}$ to draw this graph \cite{zeman1987accurate}.
  }
  \label{fig:mag.-omega}
\end{figure}

 Electrons in a metal follow the motion of the electric fields below the plasma frequency.
This implies that the circular motion of the electric field of SPP induces orbital motion of electrons in the metal,
which can be confirmed by simultaneously solving Maxwell equations and the equation of motion of electron gas in the metal \cite{bliokh2017optical_new_j_phys}.
The orbital motion of electron gas creates an inhomogeneous magnetisation field in the metal,
which is sometimes referred to as the inverse Faraday effect \cite{pitaevskii1961electric, nkoma1974elementary, kono1981spontaneous, hertel2006theory, landau2013electrodynamics}.
Using the gyromagnetic ratio for an orbiting electron \cite{herzberg1944atomic},
we can write the magnetisation density in metal
\begin{align}
  \vec{M} &= -\cfrac{e}{2mc} \vec{S}_{mat} 
  = -\cfrac{g e \omega}{4 m c} \frac{\mathrm{d}\epsilon}{\mathrm{d}\omega} \mathfrak{Im}\left( \vec{E}^* \times \vec{E} \right), 
  \notag 
  \\
          &= g |E_0|^2\frac{e}{2 mc}\frac{2(1-\epsilon)\sqrt{-\epsilon}}{\epsilon^2}e^{2\kappa_2 x} \uy 
          \label{eq:magnetisation_SPP}
          \\
          &\equiv M_0 f(\omega) e^{2\kappa_2 x} \vec{u}_y 
          \label{eq:magnetisation_SPP_def}
\end{align}
Here we set $M_0 = |E_0|^2 \frac{e}{2 mc}$ and $f(\omega) = \frac{2g(1-\epsilon)\sqrt{-\epsilon}}{\epsilon^2}$.
FIG. \ref{fig:mag.-omega} illustrates the frequency dependence of the magnetisation density.
Note that the plot is normalised by $M_{sp} = M(\omega_{sp})$.
We can find that the magnetisation density is a monotonically increasing function of the frequency,
which is maximum at the surface plasmon resonance frequency $\omega_{sp} = \omega_p/\sqrt{2}$.
From (\ref{eq:magnetisation_SPP}),
it is clear that the magnetisation density exponentially decays toward infinity in the metal.
This inhomogeneous magnetisation field could drive electron spin current.

\paragraph{Electron spin current in the inhomogeneous magnetisation field.---}
\label{para:Electron spin current---}
In order to investigate whether the inhomogeneous magnetisation of SPPs can generate electron spin currents,
we solve the spin diffusion equation \cite{valet1993theory, takahashi2008spin} with a source of the inhomogeneous magnetisation field
\begin{align}
  \left( \partial_t - D_s \nabla^2 + \frac{1}{\tau} \right) \delta \mu = \sigma_0^{-1} D_s \nabla \cdot \vec{j}_s.
  \label{eq:diffusion}
\end{align}
Here, $\delta \mu$ is the spin accumulation,
and $D_s = \lambda_s^2 / \tau$ and $\sigma_0$ are the diffusion constant and the conductivity of the metal, respectively.
The source term on the right hand side of the diffusion equation (\ref{eq:diffusion}) contains
\begin{align}
  \vec{j}_s = -\frac{\hbar \sigma_0}{m} \nabla M_y.
  \label{eq:js_source}
\end{align}
As can be seen, the source term comes from the inhomogeneous magnetisation field created by the SPP.
Our interest is to find the stationary state solution of the diffusion equation (\ref{eq:diffusion}) and to investigate whether the spin current is generated or not.
Explicitly writing the spin diffusion equation (\ref{eq:diffusion}) in the stationary state,
we obtain
\begin{align}
  \nabla^2 \delta \mu = \frac{\delta \mu}{\lambda_s^2} + \frac{\hbar}{m} \nabla^2 M_y.
  \label{eq:stationary_diffusion}
\end{align}
By solving this differential equation (\ref{eq:stationary_diffusion}), we can find that spin accumulation is created in the stationary state.
\begin{align}
  \delta \mu = \frac{\hbar M_0}{m} \frac{f(\omega)(2\kappa_2 \lambda_s)^2}{(2 \kappa_2 \lambda_s)^2 - 1} e^{2\kappa_2x}.
  \label{eq:spin_accumulation}
\end{align}

\par
\begin{figure}[htbp]
  \centering
  \includegraphics[width=\linewidth]{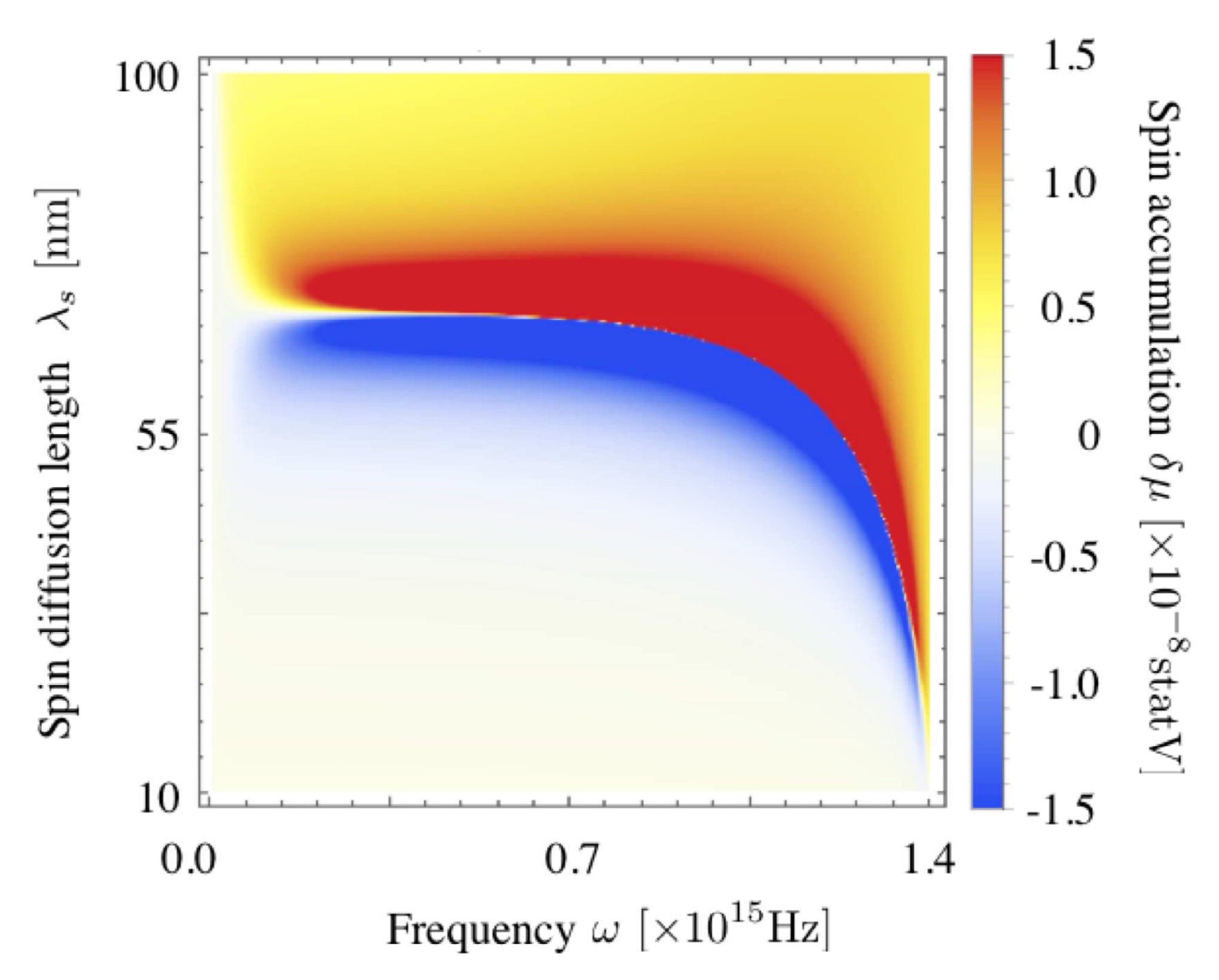}
  \caption{
    The dependence of spin accumulation on the frequency of the SPP $\omega$ and spin diffusion length $\lambda_s$.
    Note that we limit the plot range of the spin accumulation from $-1.5 \times 10^{-8}\ \mathrm{statV}$ to $1.5 \times 10^{-8}\ \mathrm{statV}$ in order to clarify the sign change at the resonant condition (\ref{eq:resonance}),
    and the accumulation takes much larger values near the condition.
    Drude parameter of gold $\omega_p = 2.15 \times 10^{15}\ \mathrm{Hz}$, $\gamma = 17.14 \times 10^{12}\ \mathrm{Hz}$ is used as in FIG. \ref{fig:mag.-omega},
    and we set $E_0 = 1.0\ \mathrm{statV/cm}$ for simplicity.
    For the spin diffusion length $\lambda_s$, we consider the range from $10\ \mathrm{nm}$ to $100\ \mathrm{nm}$,
    which is the typical range for gold \cite{takahashi2008spin}.
  }
  \label{fig:spin_acc_map}
\end{figure}

In FIG. \ref{fig:spin_acc_map},
the dependance of the spin accumulation on the SPP frequency and the spin diffusion length is shown.
We can clearly see that there is a resonant response whose condition given by
\begin{align}
  (2\kappa_2 \lambda_s)^2 - 1 & \rightarrow 0.
  \label{eq:resonance}
\end{align}
The condition is determined by the SPP frequency and the spin diffusion length of the metal.
At the condition, the sign of spin accumulation is flipped.
The accumulation takes negative values below the condition, whereas positive above it.
This Lorentz-type resonance occurs because two different parameters,
the spin diffusion length $\lambda_s$ and the decay length of SPP $\kappa_2$,
compete with each other.
Remind that the form of the stationary spin diffusion equation (\ref{eq:stationary_diffusion}) is the same as that of the differential equation of a driven harmonic oscillator.
The inverse of the spin diffusion length $(\lambda_s)^{-1}$ corresponds to the eigenfreqeuency in the harmonic oscillator equation.
These facts imply that the flow direction of the spin current generated by this SPP-induced spin accumulation can be controled by the frequency or the spin diffusion length.

\par
\begin{figure}[htbp]
  \centering
  \includegraphics[width=\linewidth]{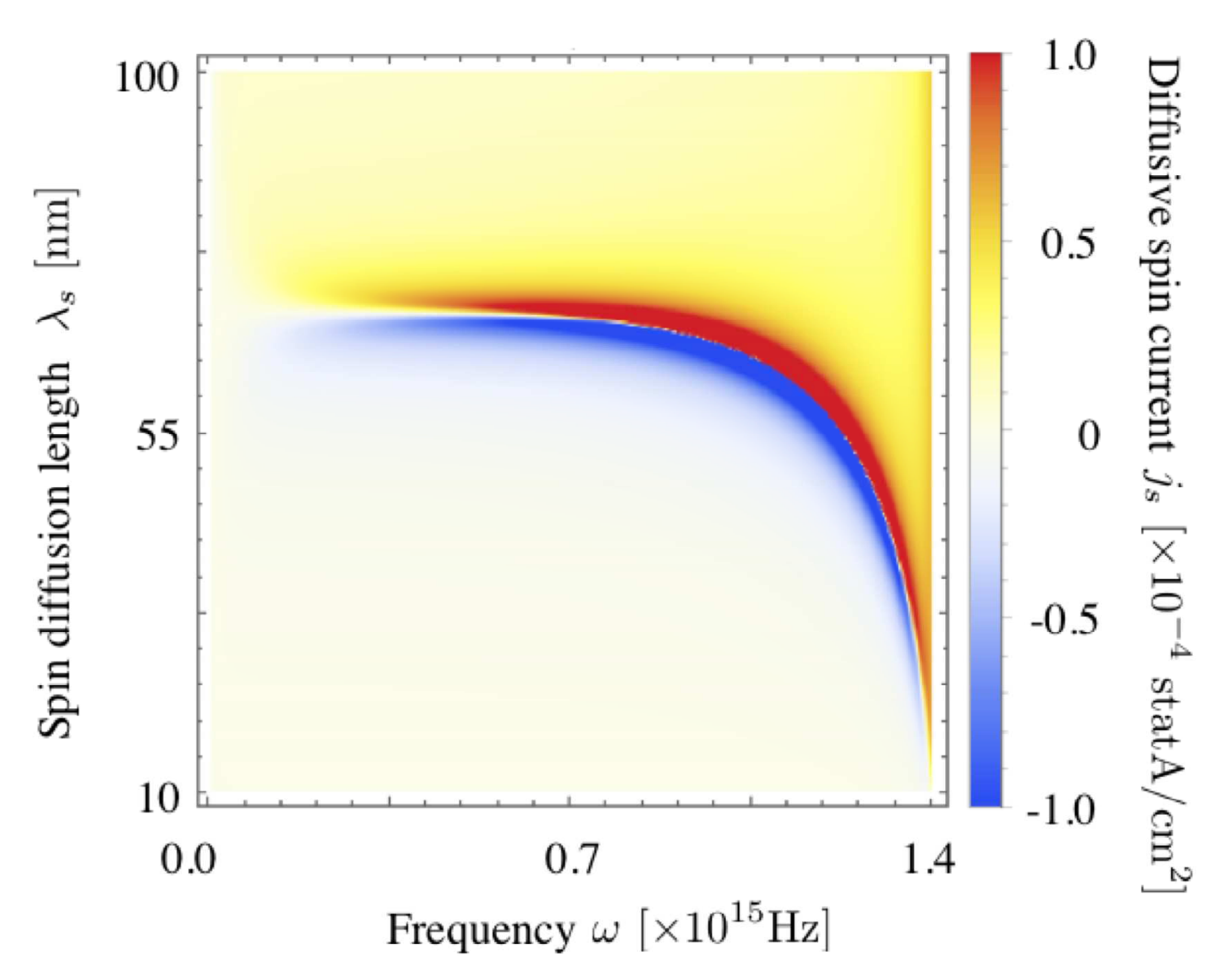}
  \caption{
    Diffusive spin current mediated by SPP.
    This colormap shows the amplitude of the spin current as a function of the frequency $\omega$ and the spin diffusion length $\lambda_s$.
    We use Drude parameter of gold as in the all figure before,
    and we also set $E_0 = 1.0\ \mathrm{statV/cm}$ for simplicity.
    It is clear that there exists the resonant response condition (\ref{eq:resonance}),
    and the flow direction of the spin current is flipped at the condition.
  Note that we set the plot range of the spin current from $-1.0 \times 10^{-4}\ \mathrm{statA/cm^2}$ to $1.0 \times 10^{-4}\ \mathrm{statA/cm^2}$,
    and that the amplitude can be larger near the resonant condition.
    We also have relatively large spin current generation near the surface plasmon resonance frequency $\omega_{sp} = \omega_p/\sqrt{2}$.
  }
  \label{fig:js^diff}
\end{figure}

Indeed, there exists the diffusive spin current driven by this spin accumulation (\ref{eq:spin_accumulation})
\begin{align}
  \vec{j}_s^{\mathrm{diff}} &= \sigma_0 \nabla \delta \mu,
  \label{eq:js^diff_def}
  \\
                            &= \frac{2 \sigma_0 \hbar M_0}{m} \frac{\kappa_2 f(\omega) (2\kappa_2 \lambda_s)^2}{(2\kappa_2 \lambda_s)^2 -1} e^{2\kappa_2 x} \vec{u}_x
                            \label{eq:js^diff_explicit}
                            \\
                            &= \frac{2 (2\kappa_2 \lambda_s)^2}{(2\kappa_2 \lambda_s)^2 -1} \vec{j}_s
                            \label{eq:js^diff}
\end{align}
whose flow direction is flipped at the resonant condition (\ref{eq:resonance}).
In FIG. \ref{fig:js^diff}, the dependence of the diffusive spin current on the SPP frequency and the spin diffusion length.
It is clear that there exists the resonant response at the condition (\ref{eq:resonance}) where the direction of the spin current is flipped.
In addition, there is another resonant response at the surface plasmon resonance frequency $\omega_{sp} = \omega_p/\sqrt{2}$ unlike the response of the spin accumulation.
This is because the decay length $\kappa_2$, which appear in (\ref{eq:js^diff}), diverges at the frequency.

\par
Finally, we estimate the amplitude of the diffusive spin current at the two resonant conditions, the surface plasmon resonance and the Lorentz-type resonance.
We here assume the electric field amplitude of SPP is $E_0 = 6.14 \times 10^2\ \mathrm{V/m}$,
which can be excited by a laser beam with an intensity of $100\ \mathrm{mW/cm^2}$ with the standard Otto configuration \cite{maier2007plasmonics}.

\par
At the surface plasmon resonance,
the magnetisation reaches its maximum of the order of $10^{-9}\ \mathrm{G} \approx 10^{-13}\ \mathrm{T}$,
and the decay length is in the order of $10^{-7}\ \mathrm{m}$.
With these values,
we can find that the amplitude of the source current (\ref{eq:js_source}) is $|\vec{j}_s| \sim 10^3\ \mathrm{A/m^{2}}$.
In the case of the surface plasmon resonance,
the Lorentz-type resonance factor $\frac{(2 \kappa_2 \lambda_s)^2}{(2 \kappa_2 \lambda_s)^2 - 1}$ is asymptotic to $1$ ($0.99$ when $\lambda_s= 40\ \mathrm{nm}$).
the corresponding diffusive spin current is in the order of $10^3\ \mathrm{A/m^{2}}$.

\par
As for the Lorentz-type resonance,
for example, when $\omega = 1.25 \times 10^{15}\ \mathrm{Hz}$ and $\lambda_s = 60\ \mathrm{nm}$,
the Lorentz-type resonance factor is of the order of $10^2$,
and we can estimate the source current $|\vec{j}_s| \sim 10^3\ \mathrm{A/m^2}$ by the same procedure as before.
Therefore, the diffusive current generated at the condition is in the order of $10^5\ \mathrm{A/m^2}$.
The spin current with an amplitude of $10^5\ \mathrm{A/m^{2}}$ can be measured via the inverse spin Hall effect (see, for example, \cite{ando2011inverse} for the ISHE measurement scheme).

\par
There may be a deviation from the simple Drude model \eqref{eq:Drude_model} due to other electronic excitations than plasma oscillation.
However, at least below the threshold energy ($ \approx 0.5\ \mathrm{PHz} \approx \omega_{sp}/3$),
the Drude model accurately fits experimental results \cite{derkachova2016dielectric},
and this allows our simple analysis by the spin diffusion equation with the source term.
We need further analysis with full quantum mechanical treatment beyond the threshold,
where the damping and the shift of the resonance peaks potentially happen,
and leave it for future work.

\paragraph{Conclusion.---}
\label{para:Conclusion---}
We reviewed the inhomogeneous magnetisation field in a surface plasmon polariton (SPP),
and proposed a mechanism of electron spin transport driven by the SPP by solving the spin diffusion equation in the presence of the inhomogeneous magnetisation of a SPP.

\par
We found that there are two conditions at which the diffusive spin current is resonantly generated.
One condition is determined by the frequency of the SPP and the spin diffusion length of electrons in the metal.
At this condition, the direction of the spin current is flipped so that we could control the direction of the electron spin flow by utilising the frequency and the spin diffusion length.
The other condition is the surface plasmon resonance (SPR) condition ($\omega = \omega_{sp}$),
where the decay length of the surface plasmon $\kappa_2$ diverges.
Unlike the former condition, 
the flip of the spin flow direction does not occur at this condition because the SPP cannot exist on the interface beyond the SPR frequency.

\par
When the electric field of the SPP is $6.14 \times 10^2\ \mathrm{V/m}$,
which can be created by a laser with a power of $100\ \mathrm{mW/cm^2}$,
the source current created by the SPP is in the order of $10^3\ \mathrm{A/m^{2}}$.
The corresponding diffusive current at the stationary state is in the order of $10^5 \mathrm{A/m^{2}}$ at one of the Lorentz-type resonance conditions ($\omega = 1.25 \times 10^{15}\ \mathrm{Hz}$ and $\lambda_s = 60\ \mathrm{nm}$),
which is measurable with the inverse spin Hall effect scheme.
Conventionally, the magneto-plasmonic effect is too weak to measure (see; for example, \cite{smolyaninov2005plasmon, bliokh2018electric});
however, the SPP-driven spin current proposed in this paper is large enough to detect via the inverse spin Hall measurement.
This is partly because of the enhancement of the spin current by the Lorentz-type resonance which results from the competition between the two parameters,
the spin diffusion length $\lambda_s$ and the decay length of the SPP $\kappa_2$,
in the spin diffusion equation with a source term.
That is also because the spin current is driven not by the magnetisation itself but by the magnetisation gradient,
which can be large since the SPP is tightly confined at the interface.
This is similar to the fact that spin currents are driven by the steep gradient of effective magnetic field created by spin-vorticity coupling in a flow of liquid metal,
although the effective magnetic field is rather weak \cite{takahashi2016spin}.

\par
Our proposed system is simple enough to prepare,
and this plasmon-mediated spin current generation will be accessible by experiments.
This work will be a bridge between two different research fields, plasmonics and spintronics.

\begin{acknowledgments}
  MM is partially Supported by the Priority Program of Chinese Academy of Sciences, Grant No. XDB28000000.
\end{acknowledgments}

\input{manuscript.bbl}

\end{document}

%% file: manuscript.bbl
%

%% file: manuscript.bbl
\begin{thebibliography}{29}%
\makeatletter
\providecommand \@ifxundefined [1]{%
 \@ifx{#1\undefined}
}%
\providecommand \@ifnum [1]{%
 \ifnum #1\expandafter \@firstoftwo
 \else \expandafter \@secondoftwo
 \fi
}%
\providecommand \@ifx [1]{%
 \ifx #1\expandafter \@firstoftwo
 \else \expandafter \@secondoftwo
 \fi
}%
\providecommand \natexlab [1]{#1}%
\providecommand \enquote  [1]{``#1''}%
\providecommand \bibnamefont  [1]{#1}%
\providecommand \bibfnamefont [1]{#1}%
\providecommand \citenamefont [1]{#1}%
\providecommand \href@noop [0]{\@secondoftwo}%
\providecommand \href [0]{\begingroup \@sanitize@url \@href}%
\providecommand \@href[1]{\@@startlink{#1}\@@href}%
\providecommand \@@href[1]{\endgroup#1\@@endlink}%
\providecommand \@sanitize@url [0]{\catcode `\\12\catcode `\$12\catcode
  `\&12\catcode `\#12\catcode `\^12\catcode `\_12\catcode `\%12\relax}%
\providecommand \@@startlink[1]{}%
\providecommand \@@endlink[0]{}%
\providecommand \url  [0]{\begingroup\@sanitize@url \@url }%
\providecommand \@url [1]{\endgroup\@href {#1}{\urlprefix }}%
\providecommand \urlprefix  [0]{URL }%
\providecommand \Eprint [0]{\href }%
\providecommand \doibase [0]{http://dx.doi.org/}%
\providecommand \selectlanguage [0]{\@gobble}%
\providecommand \bibinfo  [0]{\@secondoftwo}%
\providecommand \bibfield  [0]{\@secondoftwo}%
\providecommand \translation [1]{[#1]}%
\providecommand \BibitemOpen [0]{}%
\providecommand \bibitemStop [0]{}%
\providecommand \bibitemNoStop [0]{.\EOS\space}%
\providecommand \EOS [0]{\spacefactor3000\relax}%
\providecommand \BibitemShut  [1]{\csname bibitem#1\endcsname}%
\let\auto@bib@innerbib\@empty
\bibitem [{\citenamefont {Van~Mechelen}\ and\ \citenamefont
  {Jacob}(2016)}]{van2016universal}%
  \BibitemOpen
  \bibfield  {author} {\bibinfo {author} {\bibfnamefont {T.}~\bibnamefont
  {Van~Mechelen}}\ and\ \bibinfo {author} {\bibfnamefont {Z.}~\bibnamefont
  {Jacob}},\ }\href@noop {} {\bibfield  {journal} {\bibinfo  {journal}
  {Optica}\ }\textbf {\bibinfo {volume} {3}},\ \bibinfo {pages} {118} (\bibinfo
  {year} {2016})}\BibitemShut {NoStop}%
\bibitem [{\citenamefont {Fang}\ and\ \citenamefont
  {Wang}(2017)}]{fang2017intrinsic}%
  \BibitemOpen
  \bibfield  {author} {\bibinfo {author} {\bibfnamefont {L.}~\bibnamefont
  {Fang}}\ and\ \bibinfo {author} {\bibfnamefont {J.}~\bibnamefont {Wang}},\
  }\href@noop {} {\bibfield  {journal} {\bibinfo  {journal} {Physical Review
  A}\ }\textbf {\bibinfo {volume} {95}},\ \bibinfo {pages} {053827} (\bibinfo
  {year} {2017})}\BibitemShut {NoStop}%
\bibitem [{\citenamefont {Oue}(2019)}]{oue2019dissipation}%
  \BibitemOpen
  \bibfield  {author} {\bibinfo {author} {\bibfnamefont {D.}~\bibnamefont
  {Oue}},\ }\href@noop {} {\bibfield  {journal} {\bibinfo  {journal} {Journal
  of Optics}\ }\textbf {\bibinfo {volume} {21}},\ \bibinfo {pages} {065601}
  (\bibinfo {year} {2019})}\BibitemShut {NoStop}%
\bibitem [{\citenamefont {Novotny}\ and\ \citenamefont
  {Hecht}(2012)}]{novotny2012principles}%
  \BibitemOpen
  \bibfield  {author} {\bibinfo {author} {\bibfnamefont {L.}~\bibnamefont
  {Novotny}}\ and\ \bibinfo {author} {\bibfnamefont {B.}~\bibnamefont
  {Hecht}},\ }\href@noop {} {\emph {\bibinfo {title} {Principles of
  nano-optics}}}\ (\bibinfo  {publisher} {Cambridge university press},\
  \bibinfo {year} {2012})\BibitemShut {NoStop}%
\bibitem [{\citenamefont {Bliokh}\ and\ \citenamefont
  {Nori}(2012)}]{bliokh2012transverse}%
  \BibitemOpen
  \bibfield  {author} {\bibinfo {author} {\bibfnamefont {K.~Y.}\ \bibnamefont
  {Bliokh}}\ and\ \bibinfo {author} {\bibfnamefont {F.}~\bibnamefont {Nori}},\
  }\href@noop {} {\bibfield  {journal} {\bibinfo  {journal} {Physical Review
  A}\ }\textbf {\bibinfo {volume} {85}},\ \bibinfo {pages} {061801} (\bibinfo
  {year} {2012})}\BibitemShut {NoStop}%
\bibitem [{\citenamefont {Bliokh}\ \emph {et~al.}(2015)\citenamefont {Bliokh},
  \citenamefont {Smirnova},\ and\ \citenamefont {Nori}}]{bliokh2015quantum}%
  \BibitemOpen
  \bibfield  {author} {\bibinfo {author} {\bibfnamefont {K.~Y.}\ \bibnamefont
  {Bliokh}}, \bibinfo {author} {\bibfnamefont {D.}~\bibnamefont {Smirnova}}, \
  and\ \bibinfo {author} {\bibfnamefont {F.}~\bibnamefont {Nori}},\ }\href@noop
  {} {\bibfield  {journal} {\bibinfo  {journal} {Science}\ }\textbf {\bibinfo
  {volume} {348}},\ \bibinfo {pages} {1448} (\bibinfo {year}
  {2015})}\BibitemShut {NoStop}%
\bibitem [{\citenamefont {Bliokh}\ \emph {et~al.}(2017)\citenamefont {Bliokh},
  \citenamefont {Bekshaev},\ and\ \citenamefont
  {Nori}}]{bliokh2017optical_new_j_phys}%
  \BibitemOpen
  \bibfield  {author} {\bibinfo {author} {\bibfnamefont {K.~Y.}\ \bibnamefont
  {Bliokh}}, \bibinfo {author} {\bibfnamefont {A.~Y.}\ \bibnamefont
  {Bekshaev}}, \ and\ \bibinfo {author} {\bibfnamefont {F.}~\bibnamefont
  {Nori}},\ }\href@noop {} {\bibfield  {journal} {\bibinfo  {journal} {New
  Journal of Physics}\ }\textbf {\bibinfo {volume} {19}},\ \bibinfo {pages}
  {123014} (\bibinfo {year} {2017})}\BibitemShut {NoStop}%
\bibitem [{\citenamefont {Kato}\ \emph {et~al.}(2004)\citenamefont {Kato},
  \citenamefont {Myers}, \citenamefont {Gossard},\ and\ \citenamefont
  {Awschalom}}]{kato2004observation}%
  \BibitemOpen
  \bibfield  {author} {\bibinfo {author} {\bibfnamefont {Y.~K.}\ \bibnamefont
  {Kato}}, \bibinfo {author} {\bibfnamefont {R.~C.}\ \bibnamefont {Myers}},
  \bibinfo {author} {\bibfnamefont {A.~C.}\ \bibnamefont {Gossard}}, \ and\
  \bibinfo {author} {\bibfnamefont {D.~D.}\ \bibnamefont {Awschalom}},\
  }\href@noop {} {\bibfield  {journal} {\bibinfo  {journal} {Science}\ }\textbf
  {\bibinfo {volume} {306}},\ \bibinfo {pages} {1910} (\bibinfo {year}
  {2004})}\BibitemShut {NoStop}%
\bibitem [{\citenamefont {Wunderlich}\ \emph {et~al.}(2005)\citenamefont
  {Wunderlich}, \citenamefont {Kaestner}, \citenamefont {Sinova},\ and\
  \citenamefont {Jungwirth}}]{wunderlich2005experimental}%
  \BibitemOpen
  \bibfield  {author} {\bibinfo {author} {\bibfnamefont {J.}~\bibnamefont
  {Wunderlich}}, \bibinfo {author} {\bibfnamefont {B.}~\bibnamefont
  {Kaestner}}, \bibinfo {author} {\bibfnamefont {J.}~\bibnamefont {Sinova}}, \
  and\ \bibinfo {author} {\bibfnamefont {T.}~\bibnamefont {Jungwirth}},\
  }\href@noop {} {\bibfield  {journal} {\bibinfo  {journal} {Physical Review
  Letters}\ }\textbf {\bibinfo {volume} {94}},\ \bibinfo {pages} {047204}
  (\bibinfo {year} {2005})}\BibitemShut {NoStop}%
\bibitem [{\citenamefont {Kimura}\ \emph {et~al.}(2007)\citenamefont {Kimura},
  \citenamefont {Otani}, \citenamefont {Sato}, \citenamefont {Takahashi},\ and\
  \citenamefont {Maekawa}}]{kimura2007room}%
  \BibitemOpen
  \bibfield  {author} {\bibinfo {author} {\bibfnamefont {T.}~\bibnamefont
  {Kimura}}, \bibinfo {author} {\bibfnamefont {Y.}~\bibnamefont {Otani}},
  \bibinfo {author} {\bibfnamefont {T.}~\bibnamefont {Sato}}, \bibinfo {author}
  {\bibfnamefont {S.}~\bibnamefont {Takahashi}}, \ and\ \bibinfo {author}
  {\bibfnamefont {S.}~\bibnamefont {Maekawa}},\ }\href@noop {} {\bibfield
  {journal} {\bibinfo  {journal} {Physical Review Letters}\ }\textbf {\bibinfo
  {volume} {98}},\ \bibinfo {pages} {156601} (\bibinfo {year}
  {2007})}\BibitemShut {NoStop}%
\bibitem [{\citenamefont {Kohda}\ \emph {et~al.}(2012)\citenamefont {Kohda},
  \citenamefont {Nakamura}, \citenamefont {Nishihara}, \citenamefont
  {Kobayashi}, \citenamefont {Ono}, \citenamefont {Ohe}, \citenamefont
  {Tokura}, \citenamefont {Mineno},\ and\ \citenamefont
  {Nitta}}]{kohda2012spin}%
  \BibitemOpen
  \bibfield  {author} {\bibinfo {author} {\bibfnamefont {M.}~\bibnamefont
  {Kohda}}, \bibinfo {author} {\bibfnamefont {S.}~\bibnamefont {Nakamura}},
  \bibinfo {author} {\bibfnamefont {Y.}~\bibnamefont {Nishihara}}, \bibinfo
  {author} {\bibfnamefont {K.}~\bibnamefont {Kobayashi}}, \bibinfo {author}
  {\bibfnamefont {T.}~\bibnamefont {Ono}}, \bibinfo {author} {\bibfnamefont
  {J.-i.}\ \bibnamefont {Ohe}}, \bibinfo {author} {\bibfnamefont
  {Y.}~\bibnamefont {Tokura}}, \bibinfo {author} {\bibfnamefont
  {T.}~\bibnamefont {Mineno}}, \ and\ \bibinfo {author} {\bibfnamefont
  {J.}~\bibnamefont {Nitta}},\ }\href@noop {} {\bibfield  {journal} {\bibinfo
  {journal} {Nature Communications}\ }\textbf {\bibinfo {volume} {3}},\
  \bibinfo {pages} {1082} (\bibinfo {year} {2012})}\BibitemShut {NoStop}%
\bibitem [{\citenamefont {Takahashi}\ \emph {et~al.}(2016)\citenamefont
  {Takahashi}, \citenamefont {Matsuo}, \citenamefont {Ono}, \citenamefont
  {Harii}, \citenamefont {Chudo}, \citenamefont {Okayasu}, \citenamefont
  {Ieda}, \citenamefont {Takahashi}, \citenamefont {Maekawa},\ and\
  \citenamefont {Saitoh}}]{takahashi2016spin}%
  \BibitemOpen
  \bibfield  {author} {\bibinfo {author} {\bibfnamefont {R.}~\bibnamefont
  {Takahashi}}, \bibinfo {author} {\bibfnamefont {M.}~\bibnamefont {Matsuo}},
  \bibinfo {author} {\bibfnamefont {M.}~\bibnamefont {Ono}}, \bibinfo {author}
  {\bibfnamefont {K.}~\bibnamefont {Harii}}, \bibinfo {author} {\bibfnamefont
  {H.}~\bibnamefont {Chudo}}, \bibinfo {author} {\bibfnamefont
  {S.}~\bibnamefont {Okayasu}}, \bibinfo {author} {\bibfnamefont
  {J.}~\bibnamefont {Ieda}}, \bibinfo {author} {\bibfnamefont {S.}~\bibnamefont
  {Takahashi}}, \bibinfo {author} {\bibfnamefont {S.}~\bibnamefont {Maekawa}},
  \ and\ \bibinfo {author} {\bibfnamefont {E.}~\bibnamefont {Saitoh}},\
  }\href@noop {} {\bibfield  {journal} {\bibinfo  {journal} {Nature Physics}\
  }\textbf {\bibinfo {volume} {12}},\ \bibinfo {pages} {52} (\bibinfo {year}
  {2016})}\BibitemShut {NoStop}%
\bibitem [{\citenamefont {Kobayashi}\ \emph {et~al.}(2017)\citenamefont
  {Kobayashi}, \citenamefont {Yoshikawa}, \citenamefont {Matsuo}, \citenamefont
  {Iguchi}, \citenamefont {Maekawa}, \citenamefont {Saitoh},\ and\
  \citenamefont {Nozaki}}]{kobayashi2017spin}%
  \BibitemOpen
  \bibfield  {author} {\bibinfo {author} {\bibfnamefont {D.}~\bibnamefont
  {Kobayashi}}, \bibinfo {author} {\bibfnamefont {T.}~\bibnamefont
  {Yoshikawa}}, \bibinfo {author} {\bibfnamefont {M.}~\bibnamefont {Matsuo}},
  \bibinfo {author} {\bibfnamefont {R.}~\bibnamefont {Iguchi}}, \bibinfo
  {author} {\bibfnamefont {S.}~\bibnamefont {Maekawa}}, \bibinfo {author}
  {\bibfnamefont {E.}~\bibnamefont {Saitoh}}, \ and\ \bibinfo {author}
  {\bibfnamefont {Y.}~\bibnamefont {Nozaki}},\ }\href@noop {} {\bibfield
  {journal} {\bibinfo  {journal} {Physical Review Letters}\ }\textbf {\bibinfo
  {volume} {119}},\ \bibinfo {pages} {077202} (\bibinfo {year}
  {2017})}\BibitemShut {NoStop}%
\bibitem [{\citenamefont {Okano}\ \emph {et~al.}(2019)\citenamefont {Okano},
  \citenamefont {Matsuo}, \citenamefont {Ohnuma}, \citenamefont {Maekawa},\
  and\ \citenamefont {Nozaki}}]{okano2019nonreciprocal}%
  \BibitemOpen
  \bibfield  {author} {\bibinfo {author} {\bibfnamefont {G.}~\bibnamefont
  {Okano}}, \bibinfo {author} {\bibfnamefont {M.}~\bibnamefont {Matsuo}},
  \bibinfo {author} {\bibfnamefont {Y.}~\bibnamefont {Ohnuma}}, \bibinfo
  {author} {\bibfnamefont {S.}~\bibnamefont {Maekawa}}, \ and\ \bibinfo
  {author} {\bibfnamefont {Y.}~\bibnamefont {Nozaki}},\ }\href@noop {}
  {\bibfield  {journal} {\bibinfo  {journal} {Physical Review Letters}\
  }\textbf {\bibinfo {volume} {122}},\ \bibinfo {pages} {217701} (\bibinfo
  {year} {2019})}\BibitemShut {NoStop}%
\bibitem [{\citenamefont {Nkoma}\ \emph {et~al.}(1974)\citenamefont {Nkoma},
  \citenamefont {Loudon},\ and\ \citenamefont {Tilley}}]{nkoma1974elementary}%
  \BibitemOpen
  \bibfield  {author} {\bibinfo {author} {\bibfnamefont {J.}~\bibnamefont
  {Nkoma}}, \bibinfo {author} {\bibfnamefont {R.}~\bibnamefont {Loudon}}, \
  and\ \bibinfo {author} {\bibfnamefont {D.}~\bibnamefont {Tilley}},\
  }\href@noop {} {\bibfield  {journal} {\bibinfo  {journal} {Journal of Physics
  C: Solid State Physics}\ }\textbf {\bibinfo {volume} {7}},\ \bibinfo {pages}
  {3547} (\bibinfo {year} {1974})}\BibitemShut {NoStop}%
\bibitem [{\citenamefont {Zayats}\ \emph {et~al.}(2005)\citenamefont {Zayats},
  \citenamefont {Smolyaninov},\ and\ \citenamefont
  {Maradudin}}]{zayats2005nano}%
  \BibitemOpen
  \bibfield  {author} {\bibinfo {author} {\bibfnamefont {A.~V.}\ \bibnamefont
  {Zayats}}, \bibinfo {author} {\bibfnamefont {I.~I.}\ \bibnamefont
  {Smolyaninov}}, \ and\ \bibinfo {author} {\bibfnamefont {A.~A.}\ \bibnamefont
  {Maradudin}},\ }\href@noop {} {\bibfield  {journal} {\bibinfo  {journal}
  {Physics reports}\ }\textbf {\bibinfo {volume} {408}},\ \bibinfo {pages}
  {131} (\bibinfo {year} {2005})}\BibitemShut {NoStop}%
\bibitem [{\citenamefont {Pitaevskii}(1961)}]{pitaevskii1961electric}%
  \BibitemOpen
  \bibfield  {author} {\bibinfo {author} {\bibfnamefont {L.}~\bibnamefont
  {Pitaevskii}},\ }\href@noop {} {\bibfield  {journal} {\bibinfo  {journal}
  {Sov. Phys. JETP}\ }\textbf {\bibinfo {volume} {12}},\ \bibinfo {pages}
  {1008} (\bibinfo {year} {1961})}\BibitemShut {NoStop}%
\bibitem [{\citenamefont {Kono}\ \emph {et~al.}(1981)\citenamefont {Kono},
  \citenamefont {{\v{S}}kori{\'c}},\ and\ \citenamefont
  {Ter~Haar}}]{kono1981spontaneous}%
  \BibitemOpen
  \bibfield  {author} {\bibinfo {author} {\bibfnamefont {M.}~\bibnamefont
  {Kono}}, \bibinfo {author} {\bibfnamefont {M.}~\bibnamefont
  {{\v{S}}kori{\'c}}}, \ and\ \bibinfo {author} {\bibfnamefont
  {D.}~\bibnamefont {Ter~Haar}},\ }\href@noop {} {\bibfield  {journal}
  {\bibinfo  {journal} {Journal of Plasma Physics}\ }\textbf {\bibinfo {volume}
  {26}},\ \bibinfo {pages} {123} (\bibinfo {year} {1981})}\BibitemShut
  {NoStop}%
\bibitem [{\citenamefont {Hertel}(2006)}]{hertel2006theory}%
  \BibitemOpen
  \bibfield  {author} {\bibinfo {author} {\bibfnamefont {R.}~\bibnamefont
  {Hertel}},\ }\href@noop {} {\bibfield  {journal} {\bibinfo  {journal}
  {Journal of magnetism and magnetic materials}\ }\textbf {\bibinfo {volume}
  {303}},\ \bibinfo {pages} {L1} (\bibinfo {year} {2006})}\BibitemShut
  {NoStop}%
\bibitem [{\citenamefont {Landau}\ \emph {et~al.}(2013)\citenamefont {Landau},
  \citenamefont {Bell}, \citenamefont {Kearsley}, \citenamefont {Pitaevskii},
  \citenamefont {Lifshitz},\ and\ \citenamefont
  {Sykes}}]{landau2013electrodynamics}%
  \BibitemOpen
  \bibfield  {author} {\bibinfo {author} {\bibfnamefont {L.~D.}\ \bibnamefont
  {Landau}}, \bibinfo {author} {\bibfnamefont {J.}~\bibnamefont {Bell}},
  \bibinfo {author} {\bibfnamefont {M.}~\bibnamefont {Kearsley}}, \bibinfo
  {author} {\bibfnamefont {L.}~\bibnamefont {Pitaevskii}}, \bibinfo {author}
  {\bibfnamefont {E.}~\bibnamefont {Lifshitz}}, \ and\ \bibinfo {author}
  {\bibfnamefont {J.}~\bibnamefont {Sykes}},\ }\href@noop {} {\emph {\bibinfo
  {title} {Electrodynamics of continuous media}}},\ Vol.~\bibinfo {volume} {8}\
  (\bibinfo  {publisher} {elsevier},\ \bibinfo {year} {2013})\BibitemShut
  {NoStop}%
\bibitem [{\citenamefont {Herzberg}\ and\ \citenamefont
  {Spinks}(1944)}]{herzberg1944atomic}%
  \BibitemOpen
  \bibfield  {author} {\bibinfo {author} {\bibfnamefont {G.}~\bibnamefont
  {Herzberg}}\ and\ \bibinfo {author} {\bibfnamefont {J.~W.~T.}\ \bibnamefont
  {Spinks}},\ }\href@noop {} {\emph {\bibinfo {title} {Atomic spectra and
  atomic structure}}}\ (\bibinfo  {publisher} {Courier Corporation},\ \bibinfo
  {year} {1944})\BibitemShut {NoStop}%
\bibitem [{\citenamefont {Zeman}\ and\ \citenamefont
  {Schatz}(1987)}]{zeman1987accurate}%
  \BibitemOpen
  \bibfield  {author} {\bibinfo {author} {\bibfnamefont {E.~J.}\ \bibnamefont
  {Zeman}}\ and\ \bibinfo {author} {\bibfnamefont {G.~C.}\ \bibnamefont
  {Schatz}},\ }\href@noop {} {\bibfield  {journal} {\bibinfo  {journal}
  {Journal of Physical Chemistry}\ }\textbf {\bibinfo {volume} {91}},\ \bibinfo
  {pages} {634} (\bibinfo {year} {1987})}\BibitemShut {NoStop}%
\bibitem [{\citenamefont {Valet}\ and\ \citenamefont
  {Fert}(1993)}]{valet1993theory}%
  \BibitemOpen
  \bibfield  {author} {\bibinfo {author} {\bibfnamefont {T.}~\bibnamefont
  {Valet}}\ and\ \bibinfo {author} {\bibfnamefont {A.}~\bibnamefont {Fert}},\
  }\href@noop {} {\bibfield  {journal} {\bibinfo  {journal} {Physical Review
  B}\ }\textbf {\bibinfo {volume} {48}},\ \bibinfo {pages} {7099} (\bibinfo
  {year} {1993})}\BibitemShut {NoStop}%
\bibitem [{\citenamefont {Takahashi}\ and\ \citenamefont
  {Maekawa}(2008)}]{takahashi2008spin}%
  \BibitemOpen
  \bibfield  {author} {\bibinfo {author} {\bibfnamefont {S.}~\bibnamefont
  {Takahashi}}\ and\ \bibinfo {author} {\bibfnamefont {S.}~\bibnamefont
  {Maekawa}},\ }\href@noop {} {\bibfield  {journal} {\bibinfo  {journal}
  {Science and Technology of Advanced Materials}\ }\textbf {\bibinfo {volume}
  {9}},\ \bibinfo {pages} {014105} (\bibinfo {year} {2008})}\BibitemShut
  {NoStop}%
\bibitem [{\citenamefont {Maier}(2007)}]{maier2007plasmonics}%
  \BibitemOpen
  \bibfield  {author} {\bibinfo {author} {\bibfnamefont {S.~A.}\ \bibnamefont
  {Maier}},\ }\href@noop {} {\emph {\bibinfo {title} {Plasmonics: fundamentals
  and applications}}}\ (\bibinfo  {publisher} {Springer Science \& Business
  Media},\ \bibinfo {year} {2007})\BibitemShut {NoStop}%
\bibitem [{\citenamefont {Ando}\ \emph {et~al.}(2011)\citenamefont {Ando},
  \citenamefont {Takahashi}, \citenamefont {Ieda}, \citenamefont {Kajiwara},
  \citenamefont {Nakayama}, \citenamefont {Yoshino}, \citenamefont {Harii},
  \citenamefont {Fujikawa}, \citenamefont {Matsuo}, \citenamefont {Maekawa}
  \emph {et~al.}}]{ando2011inverse}%
  \BibitemOpen
  \bibfield  {author} {\bibinfo {author} {\bibfnamefont {K.}~\bibnamefont
  {Ando}}, \bibinfo {author} {\bibfnamefont {S.}~\bibnamefont {Takahashi}},
  \bibinfo {author} {\bibfnamefont {J.}~\bibnamefont {Ieda}}, \bibinfo {author}
  {\bibfnamefont {Y.}~\bibnamefont {Kajiwara}}, \bibinfo {author}
  {\bibfnamefont {H.}~\bibnamefont {Nakayama}}, \bibinfo {author}
  {\bibfnamefont {T.}~\bibnamefont {Yoshino}}, \bibinfo {author} {\bibfnamefont
  {K.}~\bibnamefont {Harii}}, \bibinfo {author} {\bibfnamefont
  {Y.}~\bibnamefont {Fujikawa}}, \bibinfo {author} {\bibfnamefont
  {M.}~\bibnamefont {Matsuo}}, \bibinfo {author} {\bibfnamefont
  {S.}~\bibnamefont {Maekawa}},  \emph {et~al.},\ }\href@noop {} {\bibfield
  {journal} {\bibinfo  {journal} {Journal of applied physics}\ }\textbf
  {\bibinfo {volume} {109}},\ \bibinfo {pages} {103913} (\bibinfo {year}
  {2011})}\BibitemShut {NoStop}%
\bibitem [{\citenamefont {Derkachova}\ \emph {et~al.}(2016)\citenamefont
  {Derkachova}, \citenamefont {Kolwas},\ and\ \citenamefont
  {Demchenko}}]{derkachova2016dielectric}%
  \BibitemOpen
  \bibfield  {author} {\bibinfo {author} {\bibfnamefont {A.}~\bibnamefont
  {Derkachova}}, \bibinfo {author} {\bibfnamefont {K.}~\bibnamefont {Kolwas}},
  \ and\ \bibinfo {author} {\bibfnamefont {I.}~\bibnamefont {Demchenko}},\
  }\href@noop {} {\bibfield  {journal} {\bibinfo  {journal} {Plasmonics}\
  }\textbf {\bibinfo {volume} {11}},\ \bibinfo {pages} {941} (\bibinfo {year}
  {2016})}\BibitemShut {NoStop}%
\bibitem [{\citenamefont {Smolyaninov}\ \emph {et~al.}(2005)\citenamefont
  {Smolyaninov}, \citenamefont {Davis}, \citenamefont {Smolyaninova},
  \citenamefont {Schaefer}, \citenamefont {Elliott},\ and\ \citenamefont
  {Zayats}}]{smolyaninov2005plasmon}%
  \BibitemOpen
  \bibfield  {author} {\bibinfo {author} {\bibfnamefont {I.~I.}\ \bibnamefont
  {Smolyaninov}}, \bibinfo {author} {\bibfnamefont {C.~C.}\ \bibnamefont
  {Davis}}, \bibinfo {author} {\bibfnamefont {V.~N.}\ \bibnamefont
  {Smolyaninova}}, \bibinfo {author} {\bibfnamefont {D.}~\bibnamefont
  {Schaefer}}, \bibinfo {author} {\bibfnamefont {J.}~\bibnamefont {Elliott}}, \
  and\ \bibinfo {author} {\bibfnamefont {A.~V.}\ \bibnamefont {Zayats}},\
  }\href@noop {} {\bibfield  {journal} {\bibinfo  {journal} {Physical Review
  B}\ }\textbf {\bibinfo {volume} {71}},\ \bibinfo {pages} {035425} (\bibinfo
  {year} {2005})}\BibitemShut {NoStop}%
\bibitem [{\citenamefont {Bliokh}\ \emph {et~al.}(2018)\citenamefont {Bliokh},
  \citenamefont {Rodr{\'\i}guez-Fortu{\~n}o}, \citenamefont {Bekshaev},
  \citenamefont {Kivshar},\ and\ \citenamefont {Nori}}]{bliokh2018electric}%
  \BibitemOpen
  \bibfield  {author} {\bibinfo {author} {\bibfnamefont {K.}~\bibnamefont
  {Bliokh}}, \bibinfo {author} {\bibfnamefont {F.}~\bibnamefont
  {Rodr{\'\i}guez-Fortu{\~n}o}}, \bibinfo {author} {\bibfnamefont
  {A.}~\bibnamefont {Bekshaev}}, \bibinfo {author} {\bibfnamefont
  {Y.}~\bibnamefont {Kivshar}}, \ and\ \bibinfo {author} {\bibfnamefont
  {F.}~\bibnamefont {Nori}},\ }\href@noop {} {\bibfield  {journal} {\bibinfo
  {journal} {Optics Letters}\ }\textbf {\bibinfo {volume} {43}},\ \bibinfo
  {pages} {963} (\bibinfo {year} {2018})}\BibitemShut {NoStop}%
\end{thebibliography}
